# $^{13}$C dynamic nuclear polarization in diamond via a microwave-free 'integrated' cross effect

Jacob Henshaw[1,*], Daniela Pagliero[1,*], Pablo R. Zangara[1], María B. Franzoni[5,6], Ashok Ajoy[3], Rodolfo H. Acosta[5,6], Jeffrey A. Reimer[4], Alexander Pines[3], and Carlos A. Meriles[1,2,†]

[1]Department. of Physics, CUNY-City College of New York, New York, NY 10031, USA. [2]CUNY-Graduate Center, New York, NY 10016, USA. [3]Department of Chemistry, University of California Berkeley, and Materials Science Division Lawrence Berkeley National Laboratory, Berkeley, California 94720, USA. [4]Department of Chemical and Biomolecular Engineering, and Materials Science Division Lawrence Berkeley National Laboratory University of California, Berkeley, California 94720, USA. [5]Facultad de Matemática, Astronomía, Física y Computación, Universidad Nacional de Córdoba, Ciudad Universitaria, CP:X5000HUA Córdoba, Argentina. [6]Instituto de Física Enrique Gaviola (IFEG), CONICET, Medina Allende s/n, X5000HUA, Córdoba, Argentina.

**Color-center-hosting semiconductors are emerging as promising source materials for low-field dynamic nuclear polarization (DNP) at or near room temperature, but hyperfine broadening, susceptibility to magnetic field heterogeneity, and nuclear spin relaxation induced by other paramagnetic defects set practical constraints difficult to circumvent. Here, we explore an alternate route to color-center-assisted DNP using nitrogen-vacancy (NV) centers in diamond coupled to substitutional nitrogen impurities, the so-called P1 centers. Working near the level anti-crossing condition — where the P1 Zeeman splitting matches one of the NV spin transitions — we demonstrate efficient microwave-free $^{13}$C DNP through the use of consecutive magnetic field sweeps and continuous optical excitation. The amplitude and sign of the polarization can be controlled by adjusting the low-to-high and high-to-low magnetic field sweep rates in each cycle so that one is much faster than the other. By comparing the $^{13}$C DNP response for different crystal orientations, we show that the process is robust to magnetic field/NV misalignment, a feature that makes the present technique suitable to diamond powders and settings where the field is heterogeneous. Applications to shallow NVs could capitalize on the greater physical proximity between surface paramagnetic defects and outer nuclei to efficiently polarize target samples in contact with the diamond crystal.**

Dynamic nuclear polarization | Nitrogen-vacancy centers | Substitutional nitrogen | Landau-Zener crossings

Attaining high levels of nuclear spin polarization is relevant to a broad set of applications ranging from sensitivity-enhanced nuclear magnetic resonance (NMR) spectroscopy[1] and imaging[2], to quantum simulation[3]. Over the last two decades, dynamic nuclear polarization (DNP) using molecular radicals has become ubiquitous[4], but the need for high-frequency microwave (mw) and cryogenic conditions restricts the type of samples that can be investigated, and makes the required instrumentation costly.

More recently, select color centers in wide-bandgap semiconductors such as the negatively charged nitrogen-vacancy (NV) center in diamond have been gathering growing attention as an alternative DNP platform[5-11]. Unlike most molecular radicals or inorganic paramagnetic defects, NV centers (and similar point defects in diamond[12] and other insulators[13]) can be optically spin polarized under mild optical excitation at low magnetic fields, even at room temperature. Their singular response to optical excitation along with their long spin-lattice and coherence lifetimes have already been exploited to demonstrate micro- and nano-scale chemical-shift-resolved NMR spectroscopy[14-16]. Naturally, the same properties that make these sensing schemes possible also render the NV an attractive DNP source, with recent experiments revealing polarization transfer to protons in fluids on the diamond surface[5,8]. Unfortunately, the need for paramagnetic-defect-free surfaces in properly oriented, bulk diamond crystals limit the target sample to impractically small sizes. The use of diamond powders can enormously increase the fraction of the sample in contact with the diamond surface (hence boosting the magnetization yield), but crystal-field-induced frequency shifts (e.g., from misalignment with the external magnetic field or hyperfine heterogeneity) result in broad spin resonance linewidths, difficult to manipulate efficiently[17]. Equally critical is the inevitable presence of unpolarized surface paramagnetic impurities, normally a source of nuclear spin-lattice relaxation and thus an impediment to the flow of spin polarization outside the diamond host (Fig. 1a).

This latter complication, however, could be alternatively viewed as an advantage, all the more important given the lack of NVs immediately proximal to the surface[18]. Specifically, surface paramagnetic defects interacting with deeper NVs could be exploited as proxies to mediate polarization transfer, even for quickly relaxing defects[19]. Further, many-body interactions between paramagnetic centers and nuclear spins can potentially streamline the diffusion of polarization deeper into the target at rates higher than that solely stemming from inter-nuclear dipolar couplings[20]. In this light, DNP schemes where NVs and other paramagnetic defects act cooperatively gain particular importance.

Here, we study the spin dynamics of negatively charged NV centers in diamond coupled to neighboring P1 centers: The former is a spin-1 system (with a crystal field $\Delta = 2.87$ GHz) comprising a nitrogen impurity adjacent to a vacancy, whereas the latter is a spin-1/2 point defect formed by a neutral substitutional nitrogen impurity. Assuming the presence of a magnetic field **B**, we focus on the system dynamics near the 'energy matching' condition, where the Zeeman splitting between the P1's $|m'_S = +1/2\rangle$ and $|m'_S = -1/2\rangle$ states nearly coincides with the NV $|m_S = 0\rangle \leftrightarrow |m_S = -1\rangle$ energy difference. Central to this regime is the emergence of hyperfine-shifted 'avoided energy crossings' at select magnetic fields, whose exact values depend on the NV axis orientation relative to **B**. NVs and P1s can 'cross-relax' near these fields, typically with the assistance of coupled nuclear spins, which, therefore, dynamically polarize[21,22]. Rather than operating at a constant **B**, however, the present DNP protocol uses a magnetic field of variable magnitude to sweep across the full set of avoided crossings in the presence of continuous optical excitation. Combining theory and experiment, we show the underlying DNP mechanism can be rationalized in terms of an NV–P1–$^{13}$C Landau-Zener (LZ) process featuring selective population transfer between eigenstates with opposing nuclear spin orientations. Since the transfer efficiency is a function of the sweep velocity, the sign of the nuclear spin polarization can be controlled by making the up-field or down-field segments of the cycle be faster than the other. Observations as a function of the field sweep range reveal an abrupt, sweep-direction-dependent growth of the $^{13}$C polarization, suggesting concomitant dynamic polarization of the P1 nuclear spin. Further, comparison of the bulk $^{13}$C spin signal emerging from different orientations of the diamond crystal shows the present technique is insensitive to magnetic field misalignment and thus applicable to powdered samples.

---

[*] Equally contributing authors.
[†] E-mail: cmeriles@ccny.cuny.edu

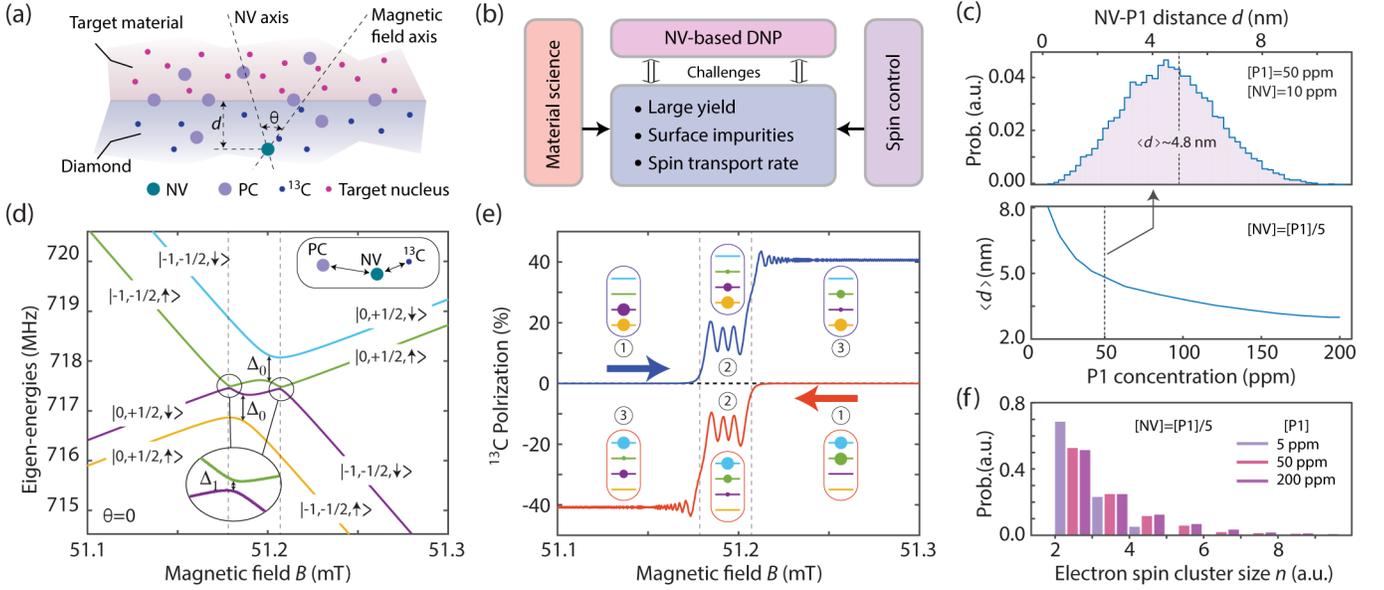

**FIG 1.** (a) Near the diamond surface, paramagnetic centers (PC) other than the NV are inevitable, though their greater abundance and proximity to the target nuclear spins can be conceivably exploited to streamline the flow of polarization away from the NV. Spin control protocols must also be robust to the misalignment θ between the magnetic field and the NV symmetry axis, unavoidable in powdered samples. (b) Advanced material science (crystal growth, ion implantation, surface termination, etc.) and spin control protocols are in demand to surmount the challenges posed by NV-based DNP. Central among them is the tradeoff between the low magnetization yield possible via bulk crystals and the high concentration of surface impurities typical in diamond powders. (c) (Top) Probability distribution of the NV–P1 distance for a P1 (NV) concentration of 50 ppm (10 ppm); the average NV-P1 distance $\langle d \rangle$ is 4.8 nm and the corresponding average NV–P1 dipolar coupling is 4 MHz (distribution not shown for brevity). (Bottom) Calculated average NV–P1 distance $\langle d \rangle$ as a function of the P1 concentration for a fixed ratio [P1]/[NV]=5. (d) Near 51 mT, the energy difference between the NV $|m_S = 0\rangle$ and $|m_S = -1\rangle$ states nearly matches the P1 Zeeman energy between states $|m'_S = +1/2\rangle$ and $|m'_S = -1/2\rangle$. The diagram shows the energy level structure around the avoided crossing assuming dipolar couplings $D_{NV-P1} = 0.5$ MHz and $D_{NV-C} = 0.92$ MHz. The upper insert shows a schematic of the model $^{13}$C–NV–P1 spin set; for simplicity, here we ignore the hyperfine couplings to the nitrogen hosts. (e) Calculated $^{13}$C polarization as a function of the magnetic field for a sweep rate of 0.26 mT/ms; we find positive (negative) polarization for a low-to-high (high-to-low) field sweep. We calculate similar polarization transfer efficiencies for the case where the $^{13}$C spin is coupled to the P1, not the NV. In (d) and (e), we assume the angle between the NV axis and the magnetic field is θ = 0. (f) Probability distribution of the electron spin cluster size $n$ for three different P1 concentrations, assuming P1s and NVs follow a 5:1 ratio. In all cases, each cluster comprises at least one NV and one P1; see text for details on the criteria for defining clusters.

**Physical principles and experimental results**

While the end use of NVs to polarize arbitrary nuclear targets will necessarily require material-science-centered sample engineering (e.g., to keep the concentration of surface defects within a manageable threshold, Fig. 1b), here we focus on the spin dynamics of carbons in the presence of energy matched NVs and P1s. Our results largely apply to other spin-1/2 defects (e.g., surface dangling bonds), and hence P1s should rather be viewed as generic, ancilla paramagnetic centers (PCs) facilitating the transfer of polarization from the NVs. Disorder in the relative positions of the electron and nuclear spins leads to a distribution of inter-defect couplings, in general a function of the NV and P1 concentrations (Fig. 1c). Fig. 1d shows a characteristic energy diagram of a $^{13}$C–NV–P1 spin set as a function of the applied magnetic field in the case where the NV axis is parallel to **B** (for simplicity, we ignore for now the NV and P1 hyperfine couplings to the $^{14}$N spins of their host nitrogens, see below). Rather than considering all possible eigen-energies, the diagram zeroes in on the $^{13}$C-spin-split branches corresponding to the electronic states $|0, +1/2\rangle$ and $|-1, -1/2\rangle$, nearly degenerate as we approach 51.2 mT. A quick inspection shows that the gap $\Delta_1$ between the inner two branches is considerably smaller than $\Delta_0$, the minimum energy difference between levels in the upper or lower pairs. Correspondingly, a suitably fast magnetic field sweep can make the ensuing LZ dynamics partly non-adiabatic only during the $\Delta_1$ crossings, without impacting the relative populations in other levels. Assuming the NV has been optically pumped into $|m_S = 0\rangle$, we conclude selective sweep-induced population exchange between the two inner branches — featuring opposite nuclear spin quantum projections — must necessarily yield a net $^{13}$C polarization (see sequence of solid circles in Fig. 1e).

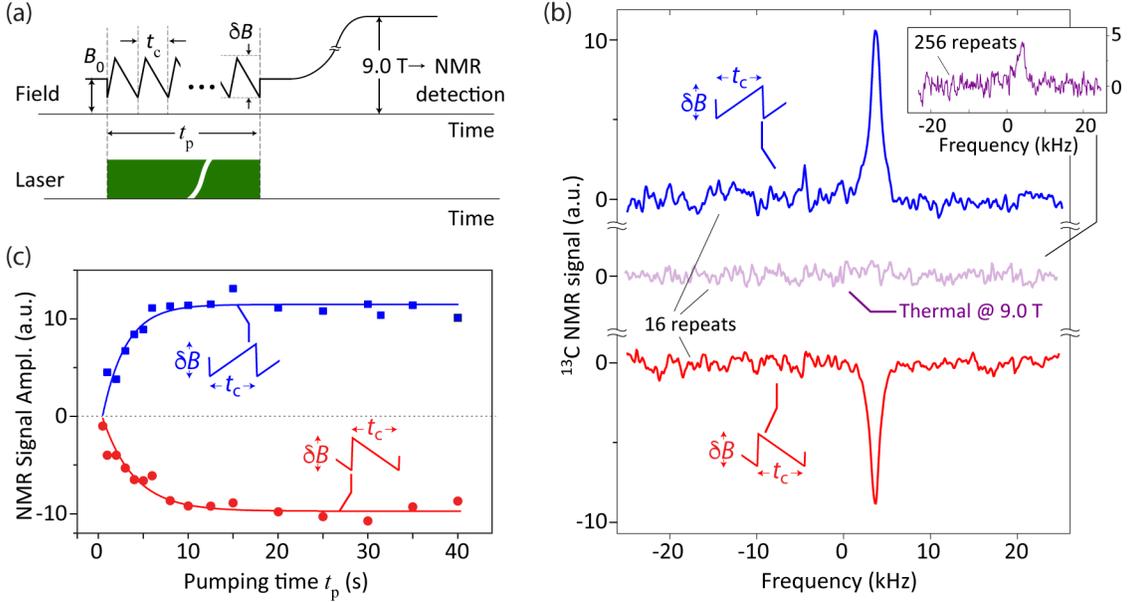

**FIG. 2.** (a) Experimental protocol. Using a shuttling device, we implement a field cycling scheme comprising optical pumping during a time $t_p$ at low magnetic field followed by $^{13}$C NMR inductive detection at 9.0 T. During optical laser excitation, we use a current amplifier to sweep the magnetic field across a range $\delta B$ centered about a static field $B_0 \sim 51$ mT. (b) Fourier transform of the $^{13}$C NMR signals after simultaneous low-field magnetic field sweeps over a range $\delta B = 6$ mT and optical excitation for a time $t_p = 10$ s. With the magnetic field sweep period unchanged ($t_c = 20.3$ ms), a positive (negative) peak emerges in the case where the low-to-high (high-to-low) segment of the sweep cycle is tenfold slower than the converse (blue and red traces, respectively). The faint, purple trace is the NMR spectrum from $^{13}$C thermal polarization at 9.0 T (no optical excitation or sample shuttling). In all three cases, the total number of repeats is 16; to account for the much longer nuclear spin-lattice relaxation time at high field, we use in the latter case a 30 min wait time between subsequent repeats. The upper right insert shows the same $^{13}$C NMR signal at high field but after 256 repeats. (c) $^{13}$C polarization buildup as a function of the pumping time $t_p$; all other conditions as in (b); solid lines indicate exponential fits. In (b) and (c), the total magnetic field is (nearly) aligned with one of the NV directions in the diamond crystal (i.e., $\theta \approx 0$). Saw-tooth traces denote field ramps where the low-to-high (blue) or the high-to-low segment (red) is ten-fold slower than the converse; $\delta B$, $t_c$ respectively denote the magnetic field range and cycle period.

In the absence of symmetry-breaking conditions, the DNP process should be identically efficient for low-to-high and high-to-low field sweeps; extending the reasoning above, however, it is not difficult to see the end polarization in either case must have opposite signs, as shown numerically in Fig. 1e. This behavior bears similarities to that observed for carbons in NV-hosting diamonds simultaneously subjected to optical excitation and mw frequency sweeps[11,23]. For future reference, we note that the spin dynamics is nearly identical in the case where the $^{13}$C spin couples to the P1, not to the NV, which should prove relevant when polarizing nuclei outside the diamond crystal[19].

Given the complexity of the interacting electron/nuclear spin ensemble, it is reasonable to question whether the three-spin set considered above can correctly capture the system's dynamics. Assuming NVs and P1s are randomly distributed throughout the diamond lattice, Fig. 1f shows the calculated electron spin cluster size distribution at two different concentrations. In our calculations, we define the cluster size $n$ as the number of electron spins whose pairwise energies are greater than $|J_d|/2$, where $|J_d|$ is the maximum NV–P1 coupling in a given set; the idea is that more-weakly coupled electron spins have little influence in the system dynamics during the LZ crossing so the boundaries between clusters of different size must be viewed as gradual. At the P1 concentration characteristic for our sample (~50 ppm, see below), we find that NV–P1 clusters ($n = 2$) are indeed the most common. The relatively higher abundance of $^{13}$C (~1%) necessarily implies that, on average, a given NV–P1 pair interacts with more than one carbon. Our calculations, however, indicate that additional nuclei do not alter the LZ spin dynamics in fundamental ways, thus validating the use of the simpler three-spin set.

To test our DNP model, we implement the protocol sketched in Fig. 2a comprising low-field optical DNP followed by detection via $^{13}$C NMR at 9 T. A full description of our experimental setup has been presented in prior work[21]. Briefly, we use a homemade shuttling mechanism to physically move the sample (a 3×3×0.3 mm$^3$ diamond crystal with natural $^{13}$C abundance, and NV and P1 concentrations of ~10 and ~50 ppm, respectively) in and out the sweet spot of the NMR magnet. To dynamically polarize the carbon spins, we expose the diamond to simultaneous laser excitation (532 nm, 500 mW) and consecutive magnetic field sweeps centered around a variable base field $B_0 \sim 46 - 56$ mT created by the proximal NMR magnet; we use an auxiliary set of coils, a programmable signal generator, and a current amplifier to induce field sweeps of variable range $\delta B$ and period $t_c$ (see Fig. 2a). In a typical run, the up-field and down-field segments of the cycle have different sweep rates, and the system undergoes multiple sweeps within the pumping time, i.e., $t_c \ll t_p$. Fig. 2b shows the Fourier transforms of the measured $^{13}$C NMR signals upon a train of sweep cycles of duration $t_c = 20.3$ ms extending over a total DNP time $t_p = 10$ s. Consistent with the model in Fig. 1, we attain positive (negative) $^{13}$C polarization when the low-to-high (high-to-low) segment of the cycle is dominant (blue and red traces in Fig. 2b, respectively). Note that although polarization of the opposite sign does build up during the shorter segment, its effect can be neglected given the much faster sweep rate (see below).

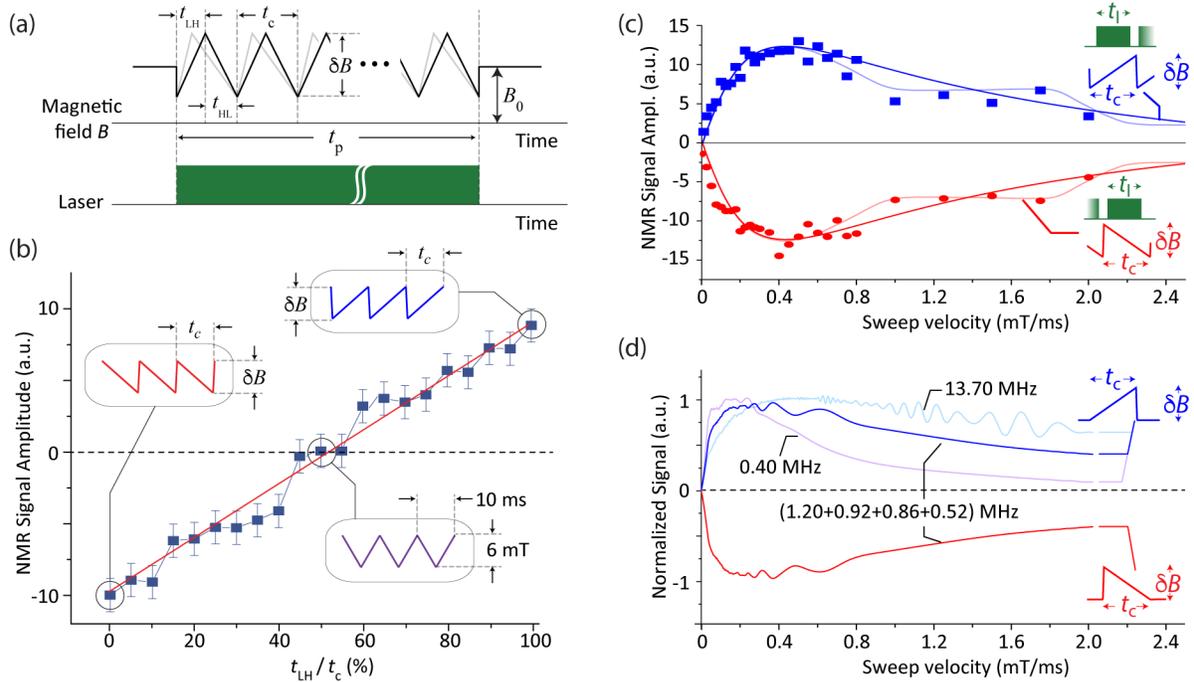

**FIG. 3.** (a) For a fixed field sweep cycle $t_c$ and field range $\delta B$, we record the $^{13}$C NMR signal (upon shuttling to 9.0 T, not shown) as we vary the ratio between the low-to-high and high-to-low intervals, respectively $t_{LH}$ and $t_{HL}$, under continuous optical excitation. (b) Amplitude of the $^{13}$C NMR signal as a function of the fractional low-to-high interval, $t_{LH}/t_c$; other conditions as in Fig. 2. The red trace is a guide to the eye. (c) $^{13}$C DNP signal amplitude as a function of the field sweep velocity; blue rectangles (red ovals) correspond to field cycles where $t_{LH}$ ($t_{HL}$) is longer. Optical excitation takes the form of a train of laser pulses, each of duration $t_l$ nearly matched to the longer segment in the field cycle (insets in the upper and lower right). Solid lines indicate fits to the expression in the text using $\Delta_1 = 30$ kHz, $\Delta_0 = 250$ kHz, $k = 15$ kHz$^2$, and $P_m = \pm 13$, with the plus and minus signs respectively corresponding to the blue and red traces. Faint solid traces are guides to eye. (d) Calculated polarization of the nuclear spin in an NV–P1–$^{13}$C trio upon a single field sweep of variable rate assuming an NV–P1 coupling of 1 MHz and different NV–$^{13}$C hyperfine couplings (faint purple and light blue traces). Best agreement with experiment is attained by adding contributions from spin trios featuring NV–$^{13}$C hyperfine couplings in the range ~1 MHz to ~0.5 MHz (red and blue traces).

Interestingly, the $^{13}$C NMR spectrum from thermal polarization at 9 T (necessarily associated to positive nuclear polarization, faint purple trace in Fig. 2b) serves as a standard to disambiguate the sign of the observed DNP signals, in principle susceptible to a uniform (but arbitrary) phase shift. Further, it allows us to estimate the NMR signal enhancement relative to 9 T, here found to be approximately thirty-fold. Since we illuminate only about ~1/50 of the crystal surface (we focus the beam to attain a saturating laser power[21]), we estimate the induced $^{13}$C polarization is of order ~1%, approximately a thousand-fold greater than the thermal value at 9 T (~10$^{-5}$). Note that because of the long $^{13}$C spin-lattice relaxation time at high field (~10 min for the present sample), a wait-time of ~30 min between successive repeats is necessary to attain representative thermal NMR signals. Therefore, even in the present conditions of limited sample excitation, the sensitivity gain due to DNP — demanding ~15 s per repeat — amounts to ~400 times.

Measurements as a function of the DNP duration $t_p$ (Fig. 2c) indicate a polarization build-up time of ~5 s, consistent with the observed $^{13}$C spin-lattice relaxation time at 50 mT[24]. Despite the slight asymmetry in the measured responses — an unwanted artifact here caused by instabilities in our home-made system — the polarization transfer efficiency in one direction or the other must be viewed as identical. We confirm this notion through the experiments in Fig. 3a, where we gradually vary the duration of the low-to-high and high-to-low segments of the sweep cycle, respectively $t_{LH}$ and $t_{HL}$, while keeping a constant period $t_c = t_{LH} + t_{HL} = 20.3$ ms. Plotting the results as a function of the fractional time $t_{LH}/t_c$ (Fig. 3b), we observe a gradual transition from negative to positive polarization, with negligible NMR signal occurring when $t_{LH} = t_{HL} = t_c/2$, i.e., precisely when positive and negative contributions to the NMR signal must counter balance.

We gain additional insight on the system spin dynamics through the observations in Fig. 3c, where we plot the measured $^{13}$C DNP amplitude as a function of the field sweep rate. To eliminate contributions of the opposite sign after a field ramp — significant at higher sweep rates given the finite reset time of **B** — we break up the optical pumping time $t_p$ into a train of pulses synchronous with the magnetic field cycle. The duration $t_l$ of each light pulse in the train is approximately matched to the longer segment in the cycle (insets in Fig. 3c). The impact of the chosen sweep rate on the observed DNP signal is major: At low sweep rates, the magnitude of the $^{13}$C polarization grows from negligible values to reach a maximum around 0.4 mT/ms; sweep rates beyond this optimum yield gradually less nuclear spin polarization.

Qualitatively, we view this response as the result of a subtle interplay between the varying degrees of adiabaticity during the spin system transition across the set of avoided crossings (Fig. 1d). Spin populations remain unchanged if the evolution is fully adiabatic, consistent with the much lower polarization buildup at low sweep rates. Conversely, exceedingly fast sweeps produce similarly non-adiabatic jumps for states with up or down nuclear spin projection, ultimately leading to poor DNP. We can crudely model the polarization transfer dynamics through the expression[11,23]

$$P(\dot{B}) = g(\dot{B})q(\dot{B})\left(1 - Q(\dot{B})\right), \quad (1)$$

where $Q(\dot{B}) = \exp(-\Delta_0^2/(|\gamma_e|\dot{B}))$ represents the probability of a transition between branches separated by the wider gap $\Delta_0$, and $\gamma_e$ denotes the electron gyromagnetic ratio (a good approximation for both the NV and P1). On the other hand, $q(\dot{B}) = \exp(-\Delta_1^2/(|\gamma_e|\dot{B}))\left(1 - Q(\dot{B})\right)$ models the transition probability across the narrower gap. Since $\Delta_1 \ll \Delta_0$, $q(\dot{B})$ quickly approaches values close to unity for moderate sweep rates but subsequently decays as the probability of a second jump to more distant branches increases. Finally, the pre-factor $g(\dot{B}) = P_m(1 - \exp(-|\gamma_e|\dot{B}/k))$ with fitting parameters $P_m$ and $k$ takes into account the varying number of field cycles during the fixed pumping time $t_p$. As shown in Fig. 3c, we attain reasonable agreement with our observations by choosing $\Delta_0 \sim 10\Delta_1 \sim 250$ kHz, consistent with the effective NV-P1 dipolar couplings present in our sample (blue and red traces). This model, however, fails to capture all observed features, particularly at higher sweep rates where the experimental response (of either sign) shows undulations absent in Eq. (1) (faint solid traces).

Fig. 3d shows the calculated nuclear polarization in an NV–P1–$^{13}$C spin trio upon a single field sweep of variable velocity. Admittedly a crude numerical representation of the dynamics at play, the results provide, nonetheless, some interesting additional clues. For example, we find that, for a given NV–P1 interaction (1 MHz in this case), strongly hyperfine-coupled carbons (~2-5 MHz and greater) are comparatively less sensitive to the sweep velocity, i.e., the polarization transfer remains efficient far above the optimum rate. Conversely, weakly coupled carbons (~0.4 MHz or less) polarize preferentially at low sweep rates, hence leading us to conclude the transfer to bulk nuclei is mostly mediated by carbons featuring intermediate hyperfine couplings (0.5-1 MHz, bright blue and red traces in Fig. 3d). In all cases, we observe smooth undulations whose amplitude and frequency tend to correlate with the strength of the NV–$^{13}$C coupling, a behavior possibly underlying the observations in Fig. 3c. Attaining a closer agreement, however, will require more involved simulations capable of taking into account spin diffusion into bulk carbons and averaging over the many different local spin configurations, a task we postpone for future studies.

Since both the NV and P1 centers feature spin active nuclear hosts, it is natural to wonder whether these interactions influence the transfer of polarization to the carbon. This question is especially pertinent in the case of the P1 center, whose nitrogen nuclear host features hyperfine couplings in excess of 100 MHz. Fig. 4a reproduces the relevant section of the energy diagram, this time adapted to a spin set where the defect nuclear hosts are both spin-1, $^{14}$N isotopes (the low natural abundance of $^{15}$N, of order 0.4%, makes other combinations highly improbable). The greater complexity of the energy level structure — apparent in the zoomed graph of Fig. 4a — is to be expected as each branch in the simpler diagram of Fig. 1d now must split into a manifold of nine levels (corresponding to spin-1 nitrogen hosts both at the P1 and NV sites). Note, however, that because the $^{14}$N coupling with the P1 is dominant (the NV–$^{14}$N interaction amounts to only ~3 MHz), branches cluster into three distinct groups, labeled in the diagram according to the quantum projection number $m'_K$ of the P1-hosting $^{14}$N.

In the simplest scenario, we may regard the nitrogen nuclear spin merely as the source of a variable local magnetic field, able to displace the energy matching condition at ~51 mT to lower or higher fields depending on the value of $m'_K$. This picture holds reasonably well in the experiments of Fig. 4b, where we measure the $^{13}$C polarization upon optical excitation at a variable magnetic field. The pattern that emerges (upper trace in Fig. 4c) can be crudely interpreted as the superposition of identical (though properly scaled) anti-symmetric motifs, each centered at a magnetic field defined by $m'_K$. In each motif, the polarization transfer dynamics effectively involves three spins (NV, P1, and $^{13}$C), with the nitrogen playing mostly a parametric role[21]. Note that although four-spin processes — including the NV–P1–$^{13}$C trio and the $^{14}$N at the P1 site — are necessary to attain full agreement[21,22], their contribution to the observed pattern is comparatively minor.

This picture breaks down in the experiments of Fig. 4d and 4e where we record the $^{13}$C DNP signal upon a field sweep of variable range; we choose the starting field to lie below ($B_0^{(u)}$) or above ($B_0^{(d)}$) the window in Fig. 4c, i.e., outside the range where optical excitation at constant **B** yields sizable $^{13}$C polarization. In the limit where the sweep range $|\delta B^{(u,d)}| \equiv |B - B_0^{(u,d)}|$ approaches the maximum, $B_0^{(d)} - B_0^{(u)}$, we attain positive or negative $^{13}$C signals, consistent with the observations in Fig. 2b. However, rather than a multi-step growth — expected if contributions from NV–P1–$^{13}$C trios with different $m'_K$ projections gradually add up — we find an abrupt transition at a magnetic field whose value depends on the direction of the sweep, namely ~49.4 mT and ~52.7 mT for the low-to-high and high-to-low ramps, respectively.

We tentatively interpret these observations as the result of a concomitant polarization of the $^{14}$N nuclear spin at the P1 site, dynamically projecting into $m'_K = \pm 1$ conditional on the sweep direction. This view provides a simple framework to understand the hysteretic behavior of Fig. 4e: If polarized at early stages of the DNP sequence, subsequent field sweep cycles merely lock the $^{14}$N spin in this projection, thus leading to a step-function-like, direction-dependent response in the $^{13}$C polarization build-up. Consistent with this notion, the 3.3 mT field span between one polarization step and the other agrees reasonably well with the field separation between the $m'_K = \pm 1$ projections in Fig. 4c. Additional work, however, will be needed to more clearly expose the spin process at play. In particular, nuclear spins coupled to the P1 rather than the NV are expected to polarize poorly when the hyperfine coupling is stronger than the Zeeman or quadrupolar interaction[21], precisely the present case. This complication may be offset here by the extreme hyperfine coupling of the P1 nitrogen (opening polarization transfer channels normally neglected) as well as its relative spatial and spectral isolation (preventing subsequent spin diffusion).

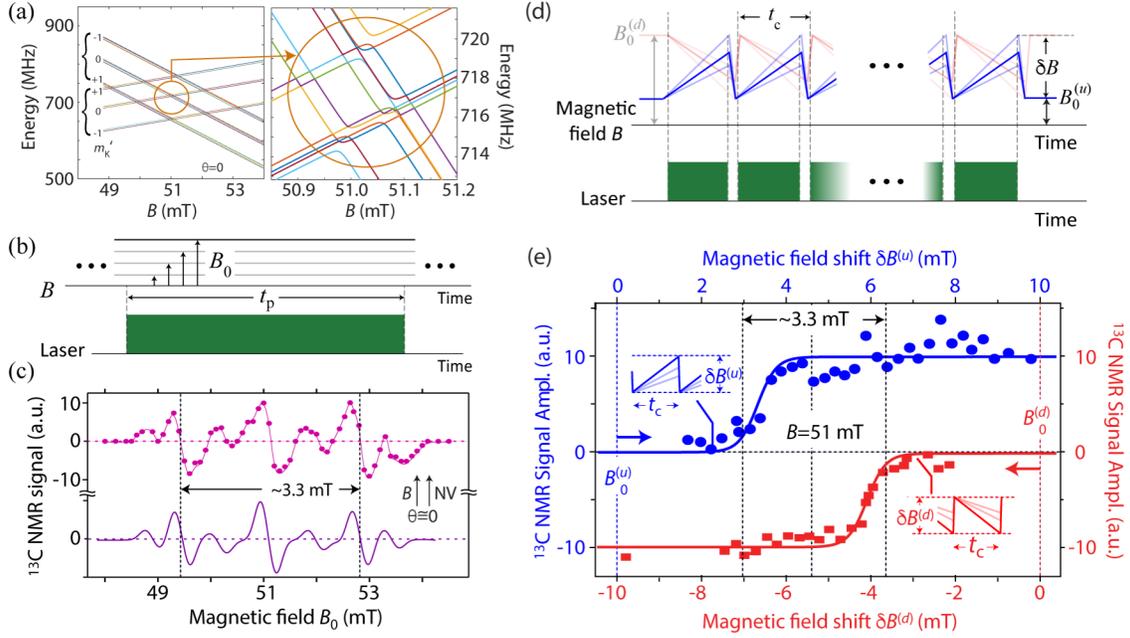

**FIG. 4.** (a) (Left) Energy diagram of an NV–P1 pair near the level (anti-)crossings. To the left of the anti-crossings, the set of states grouped by the lower (upper) bracket are of the form $|0,+1/2\rangle$ ($|-1,-1/2\rangle$); within each set, the dominant splitting stems from the P1 hyperfine interaction with the host $^{14}$N nitrogen, with quantum projection $m'_K$ along the z-axis. Additional splittings are created by the NV hyperfine interaction with its host nitrogen and nearby carbon. (Right) Zoomed level structure near the circled area assuming $D_{NV-P1} = 0.5$ MHz and $D_{NV-C} = 0.92$ MHz. (b) To determine the range of fields where DNP is active, we record the high-field $^{13}$C NMR signal upon optical excitation at an arbitrary (but fixed) magnetic field $B_0$. (c) Measured (top) and calculated (bottom) $^{13}$C NMR amplitude as a function of $B_0$ for a pumping time $t_p = 10$ s after 32 repeats; the solid trace in the upper data set is a guide to the eye. (d) (Blue trace) Starting from a low-end $B_0^{(u)} = 46$ mT, we record the $^{13}$C polarization as a function of the field sweep range, $\delta B^{(u)} = B - B_0^{(u)}$. (Faint red trace) Same as above but for a down-field sweep with origin at $B_0^{(d)} = 56$ mT and variable range $\delta B^{(d)} = B - B_0^{(d)}$. (e) $^{13}$C NMR amplitude upon implementing the protocol in (d); blue circles (red squares) emerge from the blue (faint red) sequence. All other conditions as in Fig. 2.

Interestingly, a closer inspection of Fig. 4e shows that the limit signal amplitude we attain herein is comparable to the maximum observed under static optical excitation (Fig. 4c), despite the integrative effect of the field sweep. We caution, however, that although all NV–P1 pairs participate during a sweep, the overall signal buildup depends on the hyperfine-averaged efficacy of the polarization transfer during the multiple LZ crossings, which is a function of the sweep rate and the field cycle repetition rate (itself coupled to the chosen field range). Therefore, the heuristic expectation that the signal from a field sweep must constructively add all static contributions is inaccurate in that it ignores the fundamental differences underlying the DNP dynamics in both processes.

While our experiments thus far focused on the case where the magnetic field is parallel to one of the four possible NV axis orientations ($\theta = 0$), we are now in a position to consider the more general situation. This condition is intrinsic to powders, in turn, the preferred geometry in potential applications where spins in diamond serve as the source to polarize an outside target, or when hyperpolarized particles are used as contrast agents for magnetic resonance imaging[11, 25, 26]. Fig. 5a reproduces the experiments of Fig. 4c (optically-pumped $^{13}$C NMR amplitude as a function the applied magnetic field, no sweep), but this time we include the results for a case where the magnetic field and NV axis directions do not coincide ($\theta \cong 13$ deg.). Energy matching between the NV and P1 transitions can be regained at higher magnetic fields[21], thus leading to efficient $^{13}$C DNP. The pattern of positive and negative polarization, however, moves to the right of the graph, meaning that a measurement from a sample where both orientations are present would lead, in general, to partial signal cancellation.

This problem can be circumvented with the use of a field sweep, as we show in Fig. 5b where we expand the sweep range to encompass the full field window of Fig. 5a. Remarkably, the sign of the resulting $^{13}$C DNP signal depends only on the sweep direction but not on the crystal orientation, implying that in a powdered sample all contributions would add constructively. Further, since P1-assisted DNP remains efficient for angles $\theta \approx$ 25º (or greater[21]), a moderate field sweep of just ~10 mT would be sufficient to integrate contributions within a ~50 deg. solid angle cone, a substantial fraction of the maximum possible (Figs. 5c and 5d).

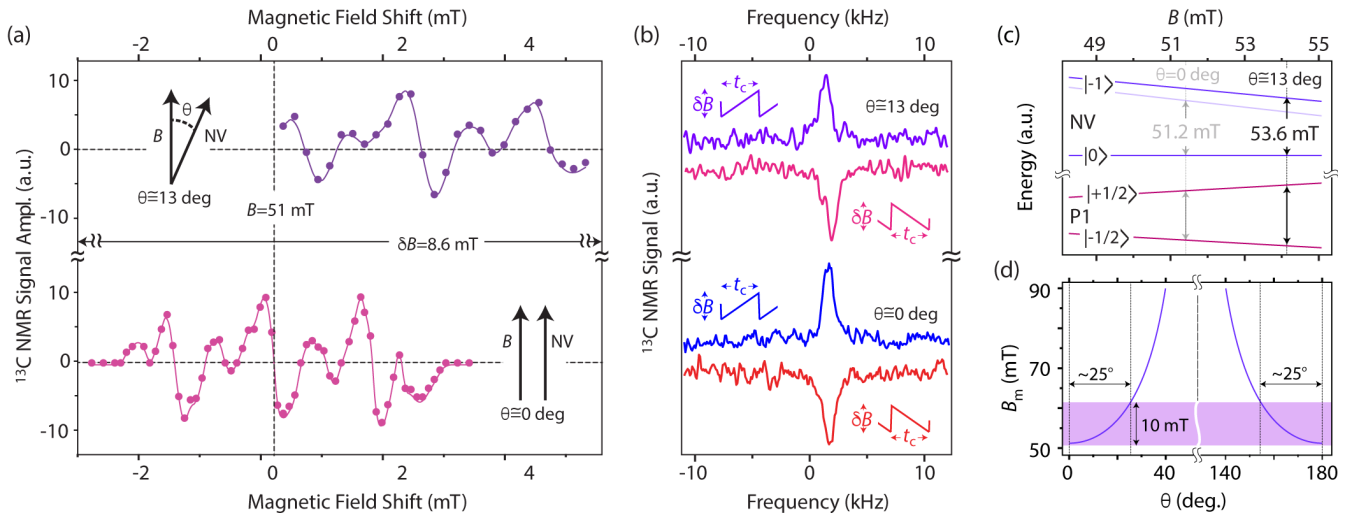

**FIG. 5.** (a) DNP spectroscopy as presented in Fig. 4b for two different orientations of the diamond crystal. Circles represent experimental data for a pumping time $t_p = 10$ s and 16 repeats; solid lines are guides to the eye. (b) $^{13}$C NMR signals upon multiple field sweeps across the range $\delta B = 8.6$ mT indicated in (a). Regardless the crystal orientation, positive (negative) $^{13}$C polarization of comparable magnitude is generated for low-to-high (high-to-low) field sweeps; spectra have been vertically displaced for clarity. In these experiments, the cycle period is $t_c = 23$ ms; all other experimental conditions as in Fig. 2. (c) Schematics of the individual NV and P1 energy level diagrams as a function of the magnetic field for $\theta = 13$ deg. When $\theta \neq 0$, matching between the $|0\rangle \leftrightarrow |-1\rangle$ and $|+1/2\rangle \leftrightarrow |-1/2\rangle$ transitions is attained at a field $B_m \geq 51$ mT (for reference, see faint energy level diagram in the back for $\theta = 0$). (d) Calculated matching field $B_m$ as a function of the relative field orientation $\theta$. Using a moderate field sweep range $\delta B \sim 10$ mT, it should be possible to integrate DNP signals within a solid cone of ~50 deg.

**Summary and outlook**

Our experiments with NV-hosting diamonds show that the combined use of optical illumination and magnetic field sweeps can yield efficient $^{13}$C dynamic nuclear polarization in the absence of microwave excitation. This process can be interpreted in terms of a partly non-adiabatic Landau-Zener population transfer near the NV-P1 level anti-crossing centered around a matching field $B_m$ where the energy separations of the NV $|0\rangle \leftrightarrow |-1\rangle$ and P1 $|+1/2\rangle \leftrightarrow |-1/2\rangle$ transitions approximately coincide. The DNP efficiency is a function of the field sweep rate, with an optimum attained near 0.4 mT/ms; for a given field cycle period and continuous optical excitation, the sign and amplitude of the nuclear spin polarization can be controlled by varying the fractional time of the up-field and down-field ramps. Observations over a variable field sweep range show a hysteretic behavior that seems to suggest sweep-direction-dependent polarization of the $^{14}$N spin at the P1 site; additional work, however, will be needed to confirm this interpretation. Experiments at different crystal orientations indicate the present DNP mechanism is relatively insensitive to the magnetic field direction and thus applicable to powdered diamond samples.

When the nuclear Zeeman (or quadrupolar) interaction is stronger than the hyperfine coupling, the energy diagram and polarization transfer mechanism of Figs. 1b and 1c remain qualitatively unchanged even when the nuclear spin couples to the P1 (or other similar paramagnetic defects) but not to the NV. Under these conditions — applicable, e.g., to protons adsorbed on the diamond surface — the present mechanism of polarization transfer can be viewed as a proxy-spin-mediated process from the NV — typically confined to comparatively deeper locations within the diamond crystal[18] — to the external nuclear target. This DNP channel should be robust to coupling strength heterogeneity, crystal orientation, and, more importantly, to proxy spin decoherence, features that could be exploited to efficiently seed spin polarization in material systems in contact with diamond[27]. The latter is consistent with detailed numerical calculations[19] showing efficient surface-defect-mediated polarization transfer to protons outside the diamond crystal and up to 6 nm from the NV (even if the proxy spin coherence time is as short as 100 ns).

Along related lines, we note that unlike other schemes relying on NV-only polarization transfer (either directly to weakly-coupled target nuclei or indirectly via spin diffusion from $^{13}$C sites), every event of NV-P1 cross relaxation is also accompanied by P1 polarization. Correspondingly, flip-flops between P1s can also lead to nuclear spin magnetization (without direct intervention of the NV). These paramagnetic-center-driven processes[20] could play a particularly important role near the diamond surface, for example, to accelerate nuclear spin-diffusion away from the diamond-target interface and deeper into the sample (on scales dictated by the inter-defect separations, not the inter-nuclear distances). Recent work in our labs on nuclear spin diffusion in diamond under NV-P1 cross relaxation conditions strongly support this notion.

Even if at the expense of greater experimental complexity, we also anticipate rotating frame, fixed-magnetic-field extensions of the present technique through the use of integrated, double-resonance DNP protocols designed to simultaneously manipulate NVs and P1s via mw frequency and amplitude modulation. Such schemes should benefit from the improved versatility of mw control, for example, to create frequency combs[28] simultaneously acting on NV–P1–$^{13}$C trios featuring different matching fields (an impossible task using the present field sweep approach). Such schemes can decouple the sweep range to be covered and the repetition rate, thus leading to more efficient DNP.

ACKNOWLEDGMENTS. J.H., D.P., P.R.Z., and C.A.M. acknowledge support from the National Science Foundation through grants NSF-1619896 and NSF-1903839, and from Research Corporation for Science Advancement through a FRED Award; they also acknowledge access to the facilities and research infrastructure of the NSF CREST IDEALS, grant number NSF-HRD-1547830. J.H. acknowledges support from CREST-PRF NSF-HRD 1827037. M.B.F. and R.H.A. acknowledge support from the Argentinean National Research Council and NSF for cooperation grant: CONICET-NSF 22620170100032 and to ANPCyT: PICT-2014-1295.

[1] D. Gajan, A. Bornet, B. Vuichoud, J. Milani, R. Melzi, H.A. van Kalkeren, L. Veyre, C. Thieuleux, M.P. Conley, W.R. Grüning, M. Schwarzwälder, A. Lesage, C. Copéret, G. Bodenhausen, L. Emsley, S. Jannin, "Hybrid polarizing solids for pure hyperpolarized liquids through dissolution dynamic nuclear polarization", *Proc. Natl. Acad. Sci. USA* **111**, 14693 (2014).

[2] R.R. Rizi, "A new direction for polarized carbon-13 MRI", *Proc. Natl. Acad. Sci. USA* **106**, 5453 (2009).

[3] J. Cai, A. Retzker, F. Jelezko, M.B. Plenio, "A large-scale quantum simulator on a diamond surface at room temperature", *Nat. Phys.* **9**, 168 (2013)

[4] T. Maly, G.T. Debelouchina, V.S. Bajaj, K.-N. Hu, C.-G. Joo, M.L. Mak–Jurkauskas, J.R. Sirigiri, P.C.A. van der Wel, J. Herzfeld, R.J. Temkin, R.G. Griffin, "Dynamic nuclear polarization at high magnetic fields", *J. Chem. Phys.* **128**, 052211 (2008).

[5] D.A. Broadway, J-P. Tetienne, A. Stacey, J.D.A.; Wood, D.A. Simpson, L.T. Hall, L.C.L. Hollenberg, "Quantum probe hyperpolarisation of molecular nuclear spins", *Nat. Commun.* **9**, 1246 (2018).

[6] J.P. King, P.J. Coles, J.A. Reimer, "Optical polarization of 13 C nuclei in diamond through nitrogen vacancy centers", *Phys. Rev. B* **81**, 073201 (2010).

[7] P. London, J. Scheuer, J.-M. Cai, I. Schwarz, A. Retzker, M.B. Plenio, M. Katagiri, T. Teraji, S. Koizumi, J. Isoya, R. Fischer, L.P. McGuinness, B. Naydenov, F. Jelezko, "Detecting and Polarizing Nuclear Spins with Double Resonance on a Single Electron Spin", *Phys. Rev. Lett.* **111**, 067601 (2013).

[8] F. Shagieva, S. Zaiser, P. Neumann, D.B.R. Dasari, R. Stöhr, A. Denisenko, R. Reuter, C.A. Meriles, J. Wrachtrup, "Microwave-assisted cross-polarization of nuclear spin ensembles from optically-pumped nitrogen-vacancy centers in diamond", *Nano Lett.* **18**, 3731 (2018).

[9] J.P. King, K. Jeong, C.C. Vassiliou, C.S. Shin, R.H. Page, C.E. Avalos, H-J. Wang, A. Pines, "Room-temperature in situ nuclear spin hyperpolarization from optically pumped nitrogen vacancy centres in diamond", *Nat. Commun.* **6**, 8965 (2015).

[10] D. Abrams, M.E. Trusheim, D. Englund, M.D. Shattuck, C.A. Meriles, "Dynamic nuclear spin polarization of liquids and gases in contact with nanostructured diamond", *Nano Lett.* **14**, 2471 (2014).

[11] A. Ajoy, K. Liu, R. Nazaryan, X. Lv, P.R. Zangara, B. Safvati, G. Wang, D. Arnold, G. Li, A. Lin, P. Raghavan, E. Druga, S. Dhomkar, D. Pagliero, J.A. Reimer, D. Suter, C.A. Meriles, A. Pines, "Orientation-independent room-temperature optical $^{13}$C hyperpolarization in powdered diamond", *Science Adv.* **4**, eaar5492 (2018).

[12] B.L. Green, B.G. Breeze, G.J. Rees, J.V. Hanna, J.-P. Chou, V. Ivády, A. Gali, M.E. Newton, "All-optical hyperpolarization of electron and nuclear spins in diamond", *Phys. Rev. B* **96**, 054101 (2017).

[13] A.L. Falk, P.V. Klimov, V. Ivády, K. Szász, D.J. Christle, W.F. Koehl, Á. Gali, D.D. Awschalom, "Optical polarization of nuclear spins in silicon carbide", *Phys. Rev. Lett.* **114**, 247603 (2015).

[14] D.R. Glenn, D.B. Bucher, J. Lee, M.D. Lukin, H. Park, R.L. Walsworth, "High-resolution magnetic resonance spectroscopy using a solid-state spin sensor", *Nature* **555**, 351 (2018).

[15] J. Smits, J. Damron, P. Kehayias, A.F. McDowell, N. Mosavian, N. Ristoff, I. Fescenko. A. Laraoui, A. Jarmola, V.M. Acosta, "Two-dimensional nuclear magnetic resonance spectroscopy with a microfluidic diamond quantum sensor", arXiv:1901.02952 (2019).

[16] N. Aslam, M. Pfender, P. Neumann, R. Reuter, A. Zappe, F. Fávaro de Oliveira, A. Denisenko, H. Sumiya, S. Onoda, J. Isoya, J. Wrachtrup, "Nanoscale nuclear magnetic resonance with chemical resolution", *Science* **357**, 67 (2017).

[17] I. Schwartz, J. Scheuer, B. Tratzmiller, S. Müller, Q. Chen, I. Dhand, Z-Y. Wang, C. Müller, B. Naydenov, F. Jelezko, M.B. Plenio, "Robust optical polarization of nuclear spin baths using Hamiltonian engineering of nitrogen-vacancy center quantum dynamics", *Sci. Adv.* **4**, eaat8978 (2018).

[18] M.V. Hauf, B. Grotz, B. Naydenov, M. Dankerl, S. Pezzagna, J. Meijer, F. Jelezko, J. Wrachtrup, M. Stutzmann, F. Reinhard, J.A. Garrido, "Chemical control of the charge state of nitrogen-vacancy centers in diamond", *Phys. Rev. B* **83**, 081304 (2011).

[19] P.R. Zangara, J. Henshaw, D. Pagliero, A. Ajoy, J.A. Reimer, A. Pines, C.A. Meriles, "Two-electron-spin ratchets as a platform for microwave-free dynamic nuclear polarization of arbitrary material targets", *Nano Lett.* **19**, 2389 (2019).

[20] J.P. Wolfe, "Direct observation of a nuclear spin diffusion barrier", *Phys. Rev. Lett.* **31**, 907 (1973).

[21] D. Pagliero, K.R. Koteswara Rao, P.R. Zangara, S. Dhomkar, H.H. Wong, A. Abril, N. Aslam, A. Parker, J. King, C.E. Avalos, A. Ajoy, J. Wrachtrup, A. Pines, C.A. Meriles, "Multispin-assisted optical pumping of bulk $^{13}$C nuclear spin polarization in diamond", *Phys. Rev. B* **97**, 024422 (2018).

[22] R. Wunderlich, J. Kohlrautz, B. Abel, J. Haase, J. Meijer, "Optically induced cross relaxation via nitrogen-related defects for bulk diamond $^{13}$C hyperpolarization", *Phys. Rev. B* **96**, 220407(R) (2017).

[23] P.R. Zangara, S. Dhomkar, A. Ajoy, K. Liu, R. Nazarian, D. Pagliero, D. Suter, J.A. Reimer, A. Pines, C.A. Meriles, "Dynamics of frequency-swept nuclear spin optical pumping in powdered diamond at low magnetic fields", *Proc. Natl. Acad. Sci. USA* **116**, 2512 (2019).

[24] A. Ajoy, B. Safvati, R. Nazaryan, J.T. Oon, B. Han, P. Raghavan, R. Nirodi, A. Aguilar, K. Liu, X. Cai, X. Lv, E. Druga, C. Ramanathan, J.A. Reimer, C.A. Meriles, D. Suter, A. Pines, "Hyperpolarized relaxometry based nuclear T1 noise spectroscopy in hybrid diamond quantum registers", *Nat. Commun.,* in press.

[25] V.N. Mochalin, O. Shenderova, D. Ho, Y. Gogotsi, "The properties and applications of nanodiamonds", *Nat. Nanotechnol.* **7,** 11 (2012).

[26] E. Rej, T. Gaebel, T. Boele, D.E.J. Waddington, D.J. Reilly, "Hyperpolarized nanodiamond with long spin-relaxation times", *Nat. Commun.* **6**, 8459 (2015).

[27] E. Rej, T. Gaebel, D.E.J. Waddington, D.J. Reilly, "Hyperpolarized nanodiamond surfaces", *J. Am. Chem. Soc.* **139**, 193 (2017).

[28] A. Ajoy, R. Nazaryan, K. Liu, X. Lv, B. Safvati, G. Wang, E. Druga, J.A. Reimer, D. Suter, C. Ramanathan, C.A. Meriles, A. Pines, "Enhanced dynamic nuclear polarization via swept microwave frequency combs", *Proc. Natl. Acad. Sci. USA* **115**, 10576 (2018).